# Uncertainty Analysis of Stray Field Measurements by Quantitative Magnetic Force Microscopy

Xiukun Hu[1], Gaoliang Dai[1], Sibylle Sievers[1], Alexander Fernández Scarioni[1], Volker Neu[2], Mark Bieler[1], and Hans Werner Schumacher[1]

[1]Physikalisch-Technische Bundesanstalt, Bundesallee 100, D-38116 Braunschweig, Germany

[2]Leibniz IFW Dresden, Helmholtzstrasse 20, 01069 Dresden, Germany

*Abstract*— **Magnetic force microscopy (MFM) measurements generally provide phase images which represent the signature of domain structures on the surface of nanomaterials. To quantitatively determine magnetic stray fields based on an MFM image requires calibrated properties of the magnetic tip. In this work, an approach is employed for calibrating a magnetic tip using a Co/Pt multilayered film as a reference sample which shows stable well-known magnetic properties and well-defined perpendicular band domains. The approach is based on a regularized deconvolution process in Fourier domain with a Wiener filter and the L-curve method for determining a suitable regularization parameter to get a physically reasonable result. The calibrated tip is applied for a traceable quantitative determination of the stray fields of a test sample which has a spatial frequency spectrum covered by that of the reference sample. According to the "Guide to the expression of uncertainty in measurement", uncertainties of the processing algorithm are estimated considering the fact that the regularization influences significantly the quantitative analysis. We discuss relevant uncertainty components and their propagations between real domain and Fourier domain for both, the tip calibration procedure and the stray field calculation, and propose an uncertainty evaluation procedure for quantitative magnetic force microscopy.**

## I. INTRODUCTION

Being a technique exploiting the interaction force between a magnetic tip and a sample, MFM has been extensively used as a powerful tool for imaging magnetic domain structure on nanometer and submicrometer scale. The technique features an excellent spatial resolution ensured by a sharp tip with a small radius down to 15 nm. With the continuous miniaturization of devices in technologically highly relevant fields such as magnetic sensors and scales, bio-medical assays, and information storage, a growing attention has been paid on a quantitative analysis of the stray fields from the phase data given directly by MFM [1-12]. To determine the stray field, the magnetic tip must be calibrated first. A few models have been discussed for determining the tip properties, e.g., by considering the magnetic tip as a point probe [10-12] and by determining its effective dipole and monopole moments from measurements above calibrated coplanar coils or hall sensors. Alternatively, tip- and lift height-dependent correlation parameters have been calibrated by measuring superparamagnetic nanoparticles [5]. However, simple point monopole and dipole models are only suitable for measurements with either very large lift height or large period of domain structures [1], or for tips with almost true point probe characteristics, such as Fe-coated Carbon needle [2]. A more general method is referred to as the tip transfer function (TTF) method [1-4, 6-10], which is based on a reference magnetic sample and is operated in Fourier domain. This tip calibration method has a few advantages over other tip calibration methods, e.g., with hall sensors or coplanar coils [10-12]: 1) It avoids the electric influence. 2) The reference sample exhibits an intrinsic, stable, periodical domain structure and generates a better-known stray field above the sample surface than the coplanar coils. 3) The domain size of the reference sample is much smaller than that of a Hall sensor, and therefore provides better calibration for quantitative high-resolution analysis. 4) Without the knowledge of the tip geometry and any assumption on its micromagnetic state, the TTF method yields the tip properties in terms of either the stray field gradient at the magnetic sample surface [6-10] or in terms of the effective magnetic charge distribution at the tip apex [1-4]. Applying the first approach, we validated the tip calibration by successfully quantifying the stray fields of a $SmCo_5$ thin film sample with perpendicular domains [9]. Furthermore, the tip calibration with the TTF method requires a deconvolution process in Fourier domain, which is an ill-posed inversion problem. Regularization methods thus need to be applied to get a reasonable estimation of the measurand. The Wiener filter [13] is normally chosen for the deconvolution process in MFM measurements [6-10]. Here, the L-curve criterion [14] is a good choice for determining the optimum inverse filter parameter.

Since the deconvolution process has significant influence on the estimation of the tip properties, and also on the determination of the stray field of test samples with the calibrated tip, an uncertainty evaluation based on the deconvolution process is necessary for quantitative MFM measurements. So far, only in [2] the authors demonstrated the uncertainty evaluation in the calibration process by averaging several instrument calibration functions in Fourier domain. A detailed uncertainty evaluation process including all input quantities has not been reported. Nowadays, the uncertainty evaluations of measurands can be done by following the "Guide to the expression of uncertainty in measurement" (GUM method) which is based on probability theory [15]. The supplement one to the GUM introduces the Monte Carlo Method (MCM) [16], which considers the propagation of probability distributions of input quantities to determine the probability density function (PDF) of the measurand and is very appropriate for complicated models. However, it is required to take normally $10^6$ trials to obtain a reasonable result, and this is very compute-intensive and time-consuming in quantitative MFM measurements, where typically more than $2.5\times10^5$ data points from a 2D MFM image need to be evaluated in each trial. The random-fuzzy variables (RFVs) method based on possibility theory [17] is an alternative approach for uncertainty evaluation. It has an advantage when handling systematic uncertainty contributions in measurements. However, since there are insignificant systematic errors in our measurements, we focus on the GUM method in this work for the sake of simplicity.

This work extends a previous conference paper [9], by considering the stray field gradient of the tip at the tip apex as the tip property, i.e., the TTF in Fourier domain. We exemplarily show a tip calibration procedure on a magnetic reference sample and present a detailed deconvolution process with the L-curve. The calibrated tip is applied for determining the stray field distributions on samples with nanoscale domain sizes. Based on the deconvolution process and the functional relationship between the measurand and input quantities, a detailed uncertainty propagation between real and Fourier domain is analyzed according to established uncertainty propagation guides [15,18-19] and the expanded uncertainty of the measurand is estimated. Our study, to the best of our knowledge, sets up for the first time a full uncertainty budget for quantitative MFM measurements based on the GUM method. It will provide valuable information both to fundamental nanomagnetic research and to industrial manufacturers for improving their quality control.

## II. Deconvolution Process

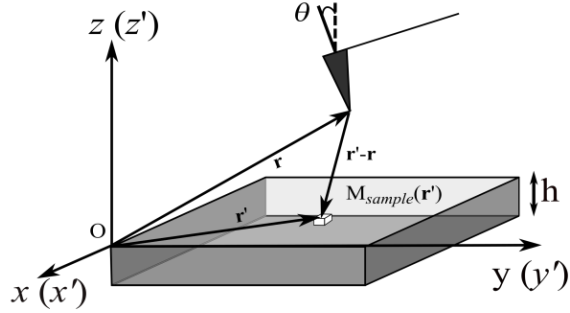

Fig. 1. Sketch of the tip-sample system and the corresponding coordinate system. The sample surface is at $z = z' = 0$. $\theta$ is the canting angle between the $z$-axis and the direction normal to the cantilever's surface. $M_{sample}(r')$ represents the local magnetization in the sample at $r' = (x', y', z')$. The tip apex locates at $r = (x, y, z)$.

The measured MFM phase shift signal reflects the force gradient from the interaction between the magnetic tip and the magnetic sample according to:

$$\Delta\phi(x,y,z) = -\frac{180}{\pi}\frac{Q}{C_{tip}}\frac{\partial F_z}{\partial z}(x,y,z), \quad (1)$$

here $\Delta\phi$ is given in units of degree, $Q$ and $C_{tip}$ are the quality factor and spring constant of the tip cantilever, respectively, $\partial F_z/\partial z (x, y, z)$ is the force gradient of the magnetic tip sensed at a distance $z$ from the sample surface, and $r = (x, y, z)$ is the position vector of the tip apex. In a tip-sample system as shown in Fig. 1, the magnetostatic energy is expressed regarding the sample magnetization in the stray field of the tip:

$$E(r) = -\mu_0 \int_{-h}^{0}\int_{-\infty}^{\infty}\int_{-\infty}^{\infty} M_{sample}(r') \cdot H_{tip}(r' - r) dx'dy'dz', \quad (2)$$

where, $M_{Sample}(r')$, $H_{tip}(r'\text{-}r)$ and $h$ are local sample magnetization, the stray field of the tip and the sample thickness, respectively. To apply the TTF method, we consider the sample as a thin film which is perpendicularly magnetized with a thickness-independent magnetization distribution $M_{sample}(r) = (0, 0, M_S m(x,y))$, where $M_S$ is the saturation magnetization, and $m(x, y)$ is the normalized perpendicular magnetization distribution. Applying the cross-correlation theorem in the $xy$ plane and referring to the derivation procedure in [1-3], we can find an expression of the force gradient acting on the tip in the discrete Fourier domain (DFT):

$$\frac{\partial F_z}{\partial z}(k, z) = \mu_0 \delta A_{pixel} M_S m(k)(1 - e^{-kh})e^{-kz}\frac{\partial \hat{H}_z^{tip}(k)}{\partial z}, \quad (3)$$

where $k = (k_x, k_y)$, $k = |k|$, is the wave vector in the two-dimensional spatial frequency domain. The additional term $\delta A_{pixel}$ is the pixel area in the discrete MFM measurement. The hat ^ refers to the complex conjugate. $\partial H_z^{tip}(k)/\partial z$ represents the partial derivative of the z component of the tip stray field with respect to $z$ in the tip apex plane. It gives an expression for calculating the stray field gradient at a position $z'$ below the tip apex as $\frac{\partial H_z^{tip}(\boldsymbol{k},z'-z)}{\partial z} = \frac{\partial H_z^{tip}(\boldsymbol{k})}{\partial z}e^{k(z'-z)}$ [3]. According to [1-3] the stray field $H_z^S$ at $z$ above the sample surface in Fourier domain can be formulated as:

$$H_z^S(\boldsymbol{k},z) = \tfrac{1}{2}M_S m(\boldsymbol{k})\left(1-e^{-kh}\right)e^{-kz}. \tag{4}$$

For the tip calibration procedure, taking into account the canting angle correction LCF($\boldsymbol{k},\theta$) of the tip cantilever as given in [2], (3) can be rephrased as:

$$\Delta\phi(\boldsymbol{k},z) = -\frac{180Q\mu_0 \delta A_{pixel} M_S}{\pi C_{tip}}[LCF(\boldsymbol{k},\theta)]^2 m(\boldsymbol{k})\left(1-e^{-kh}\right)e^{-kz}\frac{\partial \hat{H}_z^{tip}(\boldsymbol{k})}{\partial z}. \tag{5}$$

From (5), we define for the sake of simplicity:

$$G(\boldsymbol{k},z) \equiv \Delta\phi(\boldsymbol{k},z), \tag{6}$$

$$H(\boldsymbol{k},z) \equiv -[LCF(\boldsymbol{k},\theta)]^2 m(\boldsymbol{k})\left(1-e^{-kh}\right)e^{-kz}, \tag{7}$$

$$F(\boldsymbol{k},z) \equiv \frac{180Q\mu_0 \delta A_{pixel} M_S}{\pi C_{tip}}\frac{\partial \hat{H}_z^{tip}(\boldsymbol{k})}{\partial z}, \tag{8}$$

$$TTF(\boldsymbol{k},z) = \frac{\pi C_{tip}}{180Q\mu_0 \delta A_{pixel} M_S}\hat{F}(\boldsymbol{k},z), \tag{9}$$

which yields the relationship $G(\boldsymbol{k},z) = H(\boldsymbol{k},z)F(\boldsymbol{k},z)$. Here, $G$, $H$, and $F$ represent measured MFM signals, known quantities of the sample, and unknown quantities of the tip which need to be estimated, respectively. $\frac{\partial \hat{H}_z^{tip}(\boldsymbol{k})}{\partial z}$ is referred to as the tip transfer function $TTF(\boldsymbol{k},z)$. Employing the Wiener filter, $F(\boldsymbol{k},z)$ can be calculated as:

$$F(\boldsymbol{k},z) = G(\boldsymbol{k},z)\frac{\hat{H}(\boldsymbol{k},z)}{|H(\boldsymbol{k},z)|^2+\alpha}, \tag{10}$$

here $\alpha$ is a constant and referred to as the regularization parameter. In the following discussion, the three matrices $F(\boldsymbol{k},z)$, $G(\boldsymbol{k},z)$, and $H(\boldsymbol{k},z)$ are denoted as $\boldsymbol{F}$, $\boldsymbol{G}$, and $\boldsymbol{H}$, respectively. Each $\alpha$ in (10) gives a matrix $\boldsymbol{F_\alpha}$ and a residual matrix $\boldsymbol{G}-\boldsymbol{F_\alpha H}$ between the restored value $\boldsymbol{F_\alpha H}$ and the input $\boldsymbol{G}$. To obtain a physically reasonable estimation of the TTF, $\alpha$ is chosen via the L-curve criterion, in which the 2-norm $\|\boldsymbol{F_\alpha}\|_2$ is plotted as a function of the 2-norm residual $\|\boldsymbol{G}-\boldsymbol{F_\alpha H}\|_2$. The value of $\alpha$ taken from the maximal curvature of the resulting L-shape curve is selected as the optimum regularization parameter for the estimation of the TTF.

In the stray field determination procedure, (5) is rephrased by the stray field of the test sample (4), thus,

$$G(\boldsymbol{k}, z) \equiv \Delta\phi(\boldsymbol{k}, z), \tag{11}$$

$$H(\boldsymbol{k}, z) \equiv -[LCF(\boldsymbol{k}, \theta)]^2 \frac{\partial \widehat{H}_z^{tip}(\boldsymbol{k})}{\partial z}, \tag{12}$$

$$F(\boldsymbol{k}, z) \equiv \frac{360 Q \mu_0 \delta A_{pixel}}{\pi C_{tip}} H_z^S(\boldsymbol{k}, z), \tag{13}$$

$$H_z^S(\boldsymbol{k}, z) = \frac{\pi C_{tip}}{360 Q \mu_0 \delta A_{pixel}} F(\boldsymbol{k}, z). \tag{14}$$

By applying a similar deconvolution procedure as for the tip calibration, the stray fields can be quantitatively determined based on the calibrated tip, thus, being traceable to the reference sample. To apply this procedure, however, requires that the spatial frequency spectrum of the reference sample covers that of the test sample, as will be discussed later.

### III. UNCERTAINTY EVALUATION

The standard uncertainties of the input quantities $Q$, $C_{tip}$, $M_S$, $\delta A_{pixel}$, $h$ and $z$ referred to as $u_Q$, $u_{C_{tip}}$, $u_{M_S}$, $u_{\delta A_{pixel}}$, $u_h$, and $u_z$, respectively, together with their probability distributions and effective degrees of freedom can be obtained according to [15] in a straightforward way using either type A or B uncertainty evaluation. The main step for estimating the uncertainty of the measurand ($TTF$ and $H_z^S$) concerns how to obtain the uncertainty of $\boldsymbol{F}$, which involves the data transformation between the real and Fourier domains and the deconvolution process, especially on the selection of the regularization parameter in the Wiener filter. In the following subsections A-C, we first discuss the uncertainty propagation between the real and Fourier domains in the Wiener deconvolution process and estimate the combined uncertainty of $\boldsymbol{F}$. In subsections D and E, we present how to estimate the expanded uncertainty of $TTF$ and $H_z^S$ in the tip calibration and stray field determination processes, respectively.

#### A. Real Domain to Fourier Domain

For a matrix $\boldsymbol{X}$ with M×N data points and standard uncertainty $u_{X_{mn}}$ for each real value $X_{mn}$, the DFT outcome $\chi_{pq}$ can be given by

$$\chi_{pq} = \sum_{m=0}^{M-1} \sum_{n=0}^{N-1} X_{mn} e^{-i2\pi \frac{pm}{M} - i2\pi \frac{qn}{N}}, \quad \begin{array}{l} p = 0, \dots M-1 \\ q = 0, \dots N-1 \end{array}. \tag{15}$$

Extending the procedure presented in [18] from 1- to 2-dimensional data and calculating the sensitive coefficients of the input quantities $X_{mn}$, the uncertainty of $\chi_{pq}$ in Fourier domain can be represented by the real $u_{R\chi}$ and imaginary $u_{S\chi}$ parts:

$$u_{R\chi}^2 = \sum_{m=0}^{M-1} \sum_{n=0}^{N-1} \cos^2\left(2\pi \frac{pm}{M} + 2\pi \frac{qn}{N}\right) u_{X_{mn}}^2, \tag{16}$$

$$u_{S\chi}^2 = \sum_{m=0}^{M-1} \sum_{n=0}^{N-1} \sin^2\left(2\pi \frac{pm}{M} + 2\pi \frac{qn}{N}\right) u_{X_{mn}}^2, \tag{17}$$

where the subscripts $R$ and $S$ denote the real and imaginary parts, respectively.

### B. Fourier Domain to Real Domain

Similarly, the inverse DFT based on (15) results in an uncertainty propagation from Fourier domain to real domain given by:

$$u_{RX}^2 = \left(\frac{1}{MN}\right)^2 \sum_{p=0}^{M-1} \sum_{q=0}^{N-1} \left[\cos^2\left(2\pi \frac{pm}{M} + 2\pi \frac{qn}{N}\right) u_{R\chi}^2 + \sin^2\left(2\pi \frac{pm}{M} + 2\pi \frac{qn}{N}\right) u_{S\chi}^2\right]. \tag{18}$$

### C. Wiener deconvolution in Fourier Domain

Rephrasing (10) with the real and imaginary parts of $G$ and $H$ for each point as:

$$R_F = \frac{R_G R_H + S_G S_H}{R_H^2 + S_H^2 + \alpha}, \tag{19}$$

$$S_F = -\frac{R_G S_H - S_G R_H}{R_H^2 + S_H^2 + \alpha}, \tag{20}$$

where the subscripts G, H, and F denote the functions defined in (6)/(10), (7)/(11), and (8)/(12), respectively. The sensitivity coefficients $R_{RG}$, $R_{SG}$, $R_{RH}$, $R_{SH}$, $R_\alpha$, $S_{RG}$, $S_{SG}$, $S_{RH}$, $S_{SH}$, $S_\alpha$, of the input quantities $R_G$, $S_G$, $R_H$, $S_H$ and $\alpha$ can be calculated from the derivation of $R_F$ and $S_F$, e.g., $R_{RG} = \frac{\partial R_F}{\partial R_G} = \frac{R_H}{R_H^2 + S_H^2 + \alpha}$, etc., respectively. Note that $R_G$ and $S_G$ (also $R_H$ and $S_H$) are indeed correlated based on the DFT process. However, due to insufficient information for estimating the covariances, we assume here the input quantities are insignificantly correlated and the covariances are taken to be zero in this manuscript [15]. The combined uncertainty of $F$ in Fourier domain can thus be calculated as:

$$u_{RF}^2 = R_{RG}^2 u_{RG}^2 + R_{SG}^2 u_{SG}^2 + R_{RH}^2 u_{RH}^2 + R_{SH}^2 u_{SH}^2 + R_\alpha^2 u_\alpha^2, \tag{21}$$

$$u_{SF}^2 = S_{RG}^2 u_{RG}^2 + S_{SG}^2 u_{SG}^2 + S_{RH}^2 u_{RH}^2 + S_{SH}^2 u_{SH}^2 + S_\alpha^2 u_\alpha^2. \tag{22}$$

Based on above calculations and the definitions of $G$, $H$ and $F$, the combined uncertainty of $F$ in the real domain can be carried out by the following steps: 1. propagation of uncertainty of $G$ from real domain to Fourier domain using

(16) and (17); 2. obtain combined uncertainty of $H$ in Fourier domain; 3. obtain combined uncertainty of $F$ in Fourier domain using (19) - (22); 4. propagation of uncertainty of $F$ from Fourier to real domain using (18). Note that, the procedure for estimating the uncertainty of $F$ involves a huge number of independent variables, e.g., $X_{mn}$ for $G$, which are characterized by a normal distribution. The distribution of $F$ is thus approximated by a normal distribution according to the Central Limit Theorem [15].

*D. Uncertainty Propagation in the Tip calibration process*

*1) Uncertainty of random location effects*

The tip calibration procedure is performed several times (usually 5-10 times, depending on the magnetic tip) at different locations of the reference sample to obtain the mean TTF ($\overline{TTF}$). A standard deviation [15, 19] of $\overline{TTF}$ referred to as $u_{1\_TTF}$ is used as the standard uncertainty component which indicates the random location effect of the reference sample on the TTF.

*2) Uncertainty from the deconvolution process*

The tip calibration procedure invokes (6)-(10). We first need to estimate the uncertainty of $F$ in (10) in the real domain. For function $G$, the uncertainty contribution arises from $\Delta\phi$ with a standard uncertainty $u_{\Delta\phi}$ in the real domain:

$$u_G^2 = u_{\Delta\phi}^2. \tag{23}$$

The uncertainty contributions to $H$ in Fourier domain results from $h$ and $z$, (7) can be rephrased as:

$$H(\mathbf{k}, z) = (1 - e^{-kh}) \cdot e^{-kz}(R_\sigma + iS_\sigma), \tag{24}$$

$$u_{RH}^2 = k^2 R_H^2 u_z^2 + (k \cdot e^{-kh} \cdot e^{-kz})^2 R_\sigma^2 \cdot u_h^2, \tag{25}$$

$$u_{SH}^2 = k^2 S_H^2 u_z^2 + (k \cdot e^{-kh} \cdot e^{-kz})^2 S_\sigma^2 \cdot u_h^2, \tag{26}$$

Following the steps in Subsection C, the combined uncertainty $u_F$ of the function $F$ in (10) in the real domain can be estimated.

The uncertainty contributions to *TTF* from input quantities $Q$, $C_{\text{tip}}$, $M_S$, $\delta A_{\text{pixel}}$ and $F$ can thus be written as:

$$u_{2\_TTF}^2 = \overline{TTF}^2 \left[ \left(\frac{u_Q}{Q}\right)^2 + \left(\frac{u_{C_{tip}}}{C_{tip}}\right)^2 + \left(\frac{u_{M_S}}{M_S}\right)^2 + \left(\frac{u_{\delta A_{pixel}}}{\delta A_{pixel}}\right)^2 \right] + \left(\frac{\pi C_{tip}}{180 Q \mu_0 \delta A_{pixel} M_S}\right)^2 u_F^2. \tag{27}$$

*3) Combined uncertainty of TTF*

$$u_{c\_TTF}^2 = u_{1\_TTF}^2 + u_{2\_TTF}^2. \tag{28}$$

*4) Expanded uncertainty of TTF*

The effective degrees of freedom $v_{eff}$ are firstly calculated from the Welch-Satterthwaite formula as described in [15]. The probability distribution and of the coverage factor $k_{95}$ is then obtained by the *t*-factor $t_{95}(v_{eff})$ from the t-distribution for the corresponding degrees of freedom. The expanded uncertainty of *TTF* is estimated as $U_{TTF} = k_{95} u_{c\_TTF}$, providing an interval having a confidence level of 95%.

*E. Uncertainty Propagation in the Stray Field Determination with the TTF*

When the tip is calibrated with $\overline{TTF}$ and the combined uncertainty $u_{c\_TTF}$ in the real domain, it can be used to measure the stray field above a test sample and to estimate the uncertainty of the measurand. The stray field determination invokes (10)-(14). For function **G**, the uncertainty contribution arises from $\Delta\phi$ with a standard uncertainty $u_{\Delta\phi}$ in the real domain, as written in (23). The combined uncertainty of **H** in Fourier domain results from *TTF* with its uncertainties (the real part $R_{TTF}$ and $u_{R_{TTF}}$ and the imaginary part $S_{TTF}$ and $u_{S_{TTF}}$). (12) can be rephrased as:

$$H(\mathbf{k}, z) = -[R_{LCF^2} + iS_{LCF^2}] \cdot [R_{TTF} - iS_{TTF}], \qquad (29)$$

$$u_{RH}^2 = R_{LCF^2}^2 \cdot u_{R_{TTF}}^2 + S_{LCF^2}^2 \cdot u_{S_{TTF}}^2, \qquad (30)$$

$$u_{SH}^2 = S_{LCF^2}^2 \cdot u_{R_{TTF}}^2 + R_{LCF^2}^2 \cdot u_{S_{TTF}}^2. \qquad (31)$$

Following the steps in subsection C, the combined uncertainty $u_F$ of the function **F** in (10) in the real domain can be estimated.

From the function (14) the combined uncertainty of the measurand $H_z^S$ can be written as:

$$u_{c\_HZ}^2 = H_z^{S2} \left[ \left(\frac{u_Q}{Q}\right)^2 + \left(\frac{u_{C_{tip}}}{C_{tip}}\right)^2 + \left(\frac{u_{\delta A_{pixel}}}{\delta A_{pixel}}\right)^2 \right] + \left(\frac{\pi C_{tip}}{360 Q \mu_0 \delta A_{pixel}}\right)^2 u_F^2. \qquad (32)$$

Similar as in subsection D.4), the expanded uncertainty of $H_z^S$ is estimated as $U_{HZ} = k_{95} u_{c\_Hz}$ by calculating the effective degrees of freedom and *t*-factor $t_{95}(v_{eff})$ for a confidence level of 95%.

To summarize this section, (28) and (32) give the combined uncertainty of the tip transfer function *TTF* and the stray field distribution $H_z^S$ above the test sample. After calculating the coverage factor from the *t*-distribution with the effective degrees of freedom, we report the final result of the measurement as the value of the measurand with its expanded uncertainty, i.e., *TTF* with $U_{TTF}$ and $H_z^S$ with $U_{HZ}$.

IV. EXPERIMENTAL DETAILS

*A. Instrument*

MFM measurements were carried out by a metrological large-range MFM (LR-MFM) [20], which, on one hand, has resolution highly superior over other stray field detection techniques such as Hall sensors and magnetic resistance sensors, whose resolutions are limited by the device size which is larger than the radius of magnetic tips. A high precision positioning stage with position stabilities of 0.15, 0.2, and 0.6 nm in the $x$, $y$, and $z$ directions, respectively, enables LR-MFM as a powerful tool for imaging the nanoscale stray field distribution, e.g., high-density storage media and room temperature skyrmions [4, 21], where the stray fields are confined to the vicinity of the sample surface due to the small domain size. On the other hand, the large scan size up to 25 mm × 25 mm ($x$, $y$) and 5 mm ($z$) allows quantitative analysis of the stray field distribution above large-scale samples, e.g., magnetic scales. Furthermore, MFM imaging can be carried out at a certain distance away from the sample surface without calling the topography scan prior to each lift-mode scan, efficiently avoiding the wear of the tip due to rough surfaces of samples, and strongly reducing the magnetic interaction between tip and sample.

*B. Samples*

The reference sample (S0) used for tip calibration is a Co/Pt multilayered thin film with a composition of Pt(5nm)/[Pt(0.9nm) /Co(0.4nm)]$_{100}$/Pt(0.2nm) similar to the reference sample in [7-8], which exhibits a high perpendicular magnetic anisotropy, and thus shows a well-defined band domain structure with a dominant domain width of ~200 nm. The reference sample surface is relatively smooth with a *rms* roughness value of about 0.5 nm except for a few higher features with a height of up to 30 nm. The magnetic properties of the reference sample are: $M_S$ = 500 kA/m; domain wall width (DW) $\delta_{DW}$ = 16 nm; thickness of the film $h$ = 130 nm.

A patterned disk (S1) with a diameter of about 4 µm was used for the evaluation of stray field values with the calibrated tip. The film exhibits similar magnetic properties and a good overlap of the spatial frequency spectrum with the reference sample S0.

*C. MFM imaging*

Magnetic MFM tip (nanosensors PPP-MFMR) was calibrated using the sample S0 with a scan size of 5.11 µm × 5.11 µm and a pixel size of 10 nm × 10 nm. The closer to the sample surface the calibration measurement is performed, the more precise the tip information can be obtained. However, limited by the roughness of the surface and the oscillation amplitude of the tip, the lift height was chosen to be 60 nm. The same MFM measurement parameters were also applied for S1. During the measurement the tip cantilever was parallel to the *x*-axis and was tilted by 7° with respect to the sample plane.

## V. RESULT AND DISCUSSION

### A. Tip calibration

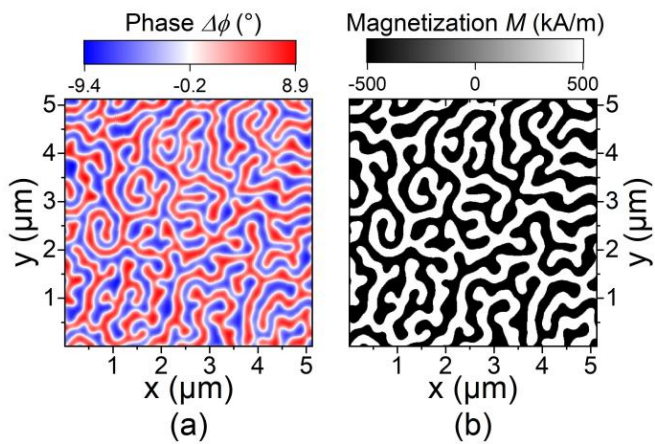

Fig. 2. (a) Typical MFM image of the reference sample S0 at $z = 60$ nm. (b) Binary image after applying a discrimination level on (a), representing the magnetization orientation in the domain structure: white: up; black: down.

Figure 2(a) shows a typical phase image of the reference sample measured with tip 1. $C_{\text{tip}}$ of the tip is given as 2.6 N/m by the manufacturer. $Q$ is 283.7 as obtained from the measured resonance curve of the tip cantilever. Before calculating the normalized perpendicular magnetization distribution $m(\mathbf{k})$, the phase image data were corrected by the canting angle function $LCF(\mathbf{k},\theta)$ in Fourier domain. After this, the magnetization distribution $m(\mathbf{k})$ on the sample top surface was obtained as a binary image by applying a discrimination level, shown in Fig. 2(b). For taking into account DW regions in which the normalized $z$ component of the magnetization follows the formula $m_z(\mathbf{r}) = \tanh(\pi r/\delta_{DW})$, a convolution operation is applied with a function $f(\mathbf{r}) = \text{sech}^2(\pi r/\delta_{DW})$. The function $H(\mathbf{k}, z)$ is then calculated by using (7).

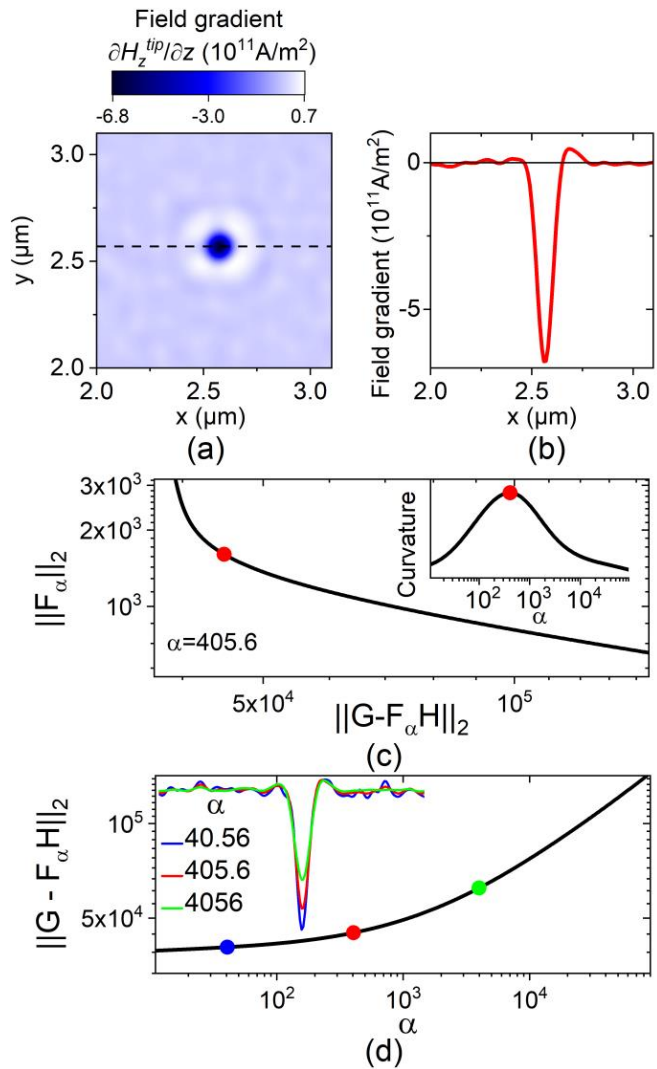

Fig. 3. (a) Calibrated field gradient $\partial H_z^{tip}/\partial z$ of the tip at the tip apex. (b) Plot of $\partial H_z^{tip}/\partial z$ profile along the dashed line indicated in (a). (c) L-curve represents the relationship between the $\alpha$-dependent 2-norm $\|\boldsymbol{F_\alpha}\|_2$ and the 2-norm $\|\boldsymbol{G} - \boldsymbol{F_\alpha H}\|_2$. The inset shows the curvature of the L-curve with respect to $\alpha$. $\alpha = 405.6$ at the peak position is the selected value for the TTF estimation indicated by red points. (d) Plot of the 2-norm $\|\boldsymbol{G} - \boldsymbol{F_\alpha H}\|_2$ as a function of $\alpha$. The inset shows the profiles $\partial H_z^{tip}/\partial z$ along the dashed line in (a) for different $\alpha$.

The calibrated $\partial H_z^{tip}/\partial z$ distribution resulting from the deconvolution process (10) is shown in Fig. 3(a) for a regularization parameter $\alpha = 405.6$. The profile along the dashed line is plotted in Fig. 3(b). The maximum value is about $6.78 \times 10^{11}$ A/m$^2$. Non-vanishing values occur in a region around the tip apex with a radius ~ 200 nm, resulting from the limited range of domain sizes of the reference sample. The L-curve during the deconvolution process is shown in Fig. 3(c). The $\alpha$ was taken from the peak position in the plot of the L-curve curvature as a function of $\alpha$, as shown in the inset of Fig. 3(c). Furthermore, the 2-norm residual between the restored image and input phase image is also plotted as a function of $\alpha$ in Fig. 3(d). The best $\alpha$ is denoted as red solid points. This curve shows that the residual approaches zero with decreasing $\alpha$ to 0. However, a too small $\alpha$ leads to physically unreasonable $\partial H_z^{tip}/\partial z$ profiles with large noise levels while a too large $\alpha$ strongly reduces the $\partial H_z^{tip}/\partial z$ value. This can be seen from the insets in Fig. 3(d), where the line profiles across the tip center are shown for different $\alpha$. In addition, the selected $\alpha$ also corresponds to the maximal curvature of the $\|G - F_\alpha H\|_2$ versus $\alpha$ curve. Therefore, we believe that the selection of $\alpha$ with the L-curve method gives the best estimation of the TTF. The mean residual value between the restored image and the input image is about 0.15° and is comparable to the standard uncertainty of noise (0.22°) in this measurement.

The above procedure was performed for 7 individual MFM images taken at different locations on the reference sample. The mean TTF is shown in Fig. 4(a). Compared with a single scan, e.g. Fig. 3(a), the mean TTF shows depressed noise signals outside the effective area of the tip (500 nm in diameter). The uncertainty evaluation was performed according to the subsection D. All standard uncertainties of the input quantities together with their distribution and degrees of freedom obtained from a Type A or estimated from a Type B evaluation for the tip calibration are listed in the Table I. $\partial H_z^{tip}/\partial z$ (black lines which cross the center) and its different types of uncertainties

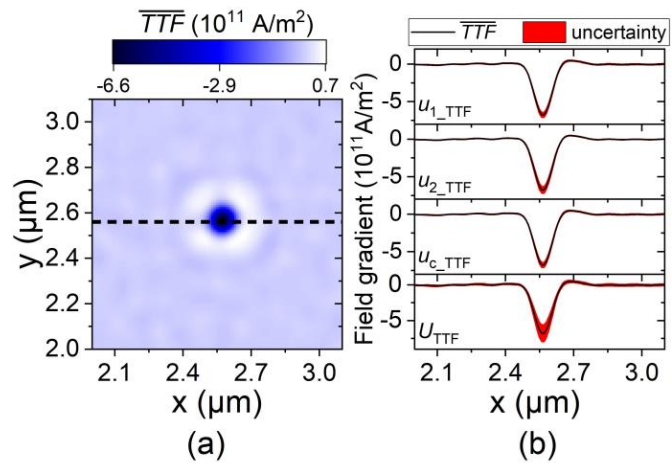

Fig. 4. (a) Mean *TTF* of 7 times MFM. (b) $\overline{TTF}$ and its uncertainty profiles along the dashed line indicated in Fig. 4(a). (From top to bottom: $u_{1\_TTF}$, $u_{2\_TTF}$, the combined uncertainties $u_{c\_TTF}$ and the expanded uncertainty $U_{TTF}$.)

(red shadows) are shown in Fig. 4 (b). For the maximal value of the TTF, $u_{1\_TTF}$, $u_{2\_TTF}$, and $u_{c\_TTF}$ are $0.32 \times 10^{11}$, $0.44 \times 10^{11}$, $0.54 \times 10^{11}$ A/m² with $v_{eff}$ of 6, 8, and 13, respectively. The expanded uncertainty for the maximal value is thus calculated as $U_{TTF} = 1.17 \times 10^{11}$ A/m² ($k_{95}=2.16$). The expanded uncertainty is about 16.2% for the maximal value of the *TTF*. Analyzing $u_{2\_TTF}$, the uncertainty components of the input quantities $Q$, $C_{tip}$, $M_S$, $\delta A_{pixel}$ are dominant, while the contribution from the deconvolution process, i.e., $F$ in (27), is one order smaller than the other uncertainty components. e.g., $0.02\times10^{11}$ A/m$^2$ for the maximal value. This might be due to the small uncertainty of $\alpha$, indicating that an optimization of the deconvolution may significantly reduce the uncertainty. Furthermore, the effect of the correlations between input quantities in the deconvolution process on the uncertainty estimation is still unclear. It is worth noting that the MCM could be employed to study this effect by considering the PDF of the input quantities. However, due to the complicated model for the TTF calibration based on a 2D image with large amount of input variables, the implementation of MCM takes large computing time and computer memory in a typical simulation with $10^6$ trials. Therefore, we postpone this issue and a comparison between GUM and MCM to later studies.

TABLE I
UNCERTAINTY COMPONENTS FOR TIP CALIBRATION

| Type | Uncertainty component | | | value | Standard uncertainty | Distribution[*] | Degrees of freedom |
|---|---|---|---|---|---|---|---|
| A | Random location effects (A/m$^2$) | | | $\overline{TTF}$ | $u_{1\_TTF}$ | t | 6 |
|  | $Q$ | | | 283.7 | 6.1 | t | 78 |
|  | $M_S$ (kA/m) | | | 500 | 25 | t | 3 |
| B | F | G | $\Delta\phi$ (°) | $\Delta\phi$ | 0.22 | N | $\infty$ |
|  |  | H | $h$ (nm) | 130 | 5 | Rec (15) | $\infty$ |
|  |  |  | $z$ (nm) | 60 | 1 | Rec (3) | $\infty$ |
|  |  | log$\alpha$ | log$\alpha$ | 2.61 | 0.09 | Rec (0.3) | $\infty$ |
|  | $C_{tip}$ (N/m) | | | 2.60 | 0.06 | Rec (0.2) | $\infty$ |
|  | $\delta A_{pixel}$ (nm$^2$) | | | 100.0 | 2.5 | N | $\infty$ |

[*] t: t-distribution; N: Normal distribution; Rec(A): rectangular distribution. The number A in brackets indicates the full width of the distribution.

The calibrated tip was used for an MFM measurement on the test sample S1 to evaluate the stray fields. Fig. 5(a) shows the MFM image of S1 measured by the calibrated tip at a lift height $z = 60$ nm. Using (10) - (14), the corresponding stray field distribution at a distance of 60 nm from the sample surface is evaluated, as shown in Fig. 5(b). Fig. 5(c) shows the profiles of the stray field along the line in Fig. 5(b) obtained from the calibrated tip (black) and from calculation using (4) (green). The stray field profiles match well with each other except that the green one shows higher values. It might be due to the higher magnetic moment of S1 (1.7 times higher) than S0, probably leading to a magnetization alignment not fully out-of-plane [8]. The stray field calculated based on (4) is then overestimated.

Furthermore, the roughness of the sample after the patterning process might also influences the MFM phase signal, giving rise to the deviations.

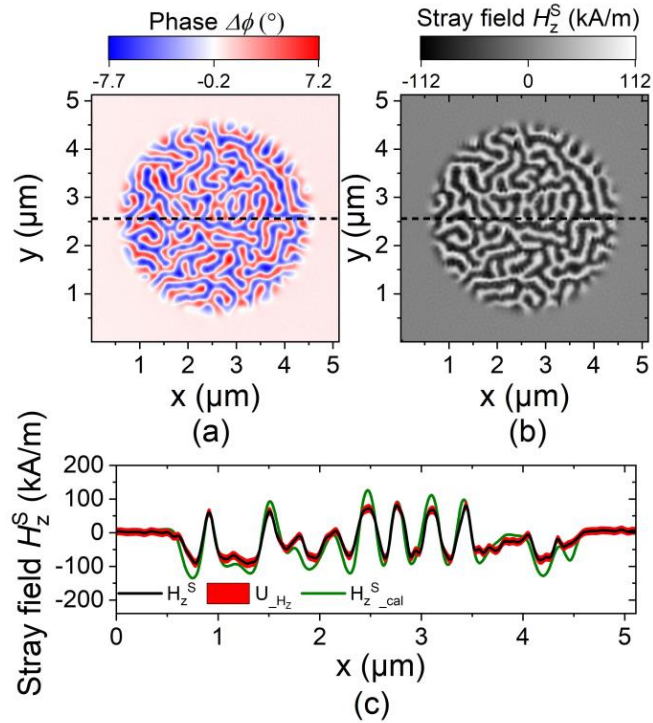

Fig. 5. (a) MFM image of the patterned sample S1 at $z = 60$ nm with the calibrated tip. (b). Determined stray field distribution at $z = 60$ nm by the calibrated tip. (c) Stray field profiles taken from the line in (b) (black) and calculated curve using (4) (green). Expanded uncertainties of the stray field from the deconvolution procedure are shown as red shadow.

TABLE II
UNCERTAINTY COMPONENT FOR STRAY FIELD DETERMINATION

| Uncertainty component | | | value | Standard uncertainty | Distribution | Degrees of freedom |
|---|---|---|---|---|---|---|
| $F$ | $G$ | $\Delta\phi$ (°) | $\Delta\phi$ | 0.11 | N | ∞ |
| | $H$ | $TTF$ (A/m$^2$) | $\overline{TTF}$ | $u_{c\_TTF}$ | $t_v(\overline{TTF}, u^2_{c\_TTF})$ | $v_{\text{eff\_TTF}}$ |
| | $\log\alpha$ | $\log\alpha$ | 25.49 | $\log\alpha$=0.12 | Rec (0.4) | ∞ |
| $Q$ $C_{\text{tip}}$ (N/m) $\delta A_{\text{pixel}}$ (nm$^2$) | | | | Listed in TABLE I | | |

The uncertainty evaluation procedure following (29) - (32) was also carried out for this measurement. All standard uncertainties of the input quantities in this process are listed in the Table II. For the absolute maximal stray field value of 111.6 kA/m, the contribution from the deconvolution process, i.e., $TTF$, to $u_{c\_HZ}$ is about 5.3 kA/m while 4.5 kA/m from other input quantities i.e., $Q$, $C_{\text{tip}}$, $\delta A_{\text{pixel}}$. As mentioned in the Section III, $u_F$ in (32) can be approximated with a normal distribution. For the calculation of $v_{\text{eff}}$ only the contribution from $Q$ is important because the degrees of freedom of the other input quantities are infinity. $v_{\text{eff}}$ is estimated as 5488. Hence, $H_z^S$ is also approximated with a normal distribution. The coverage factor can be taken as $k_{95}=2$ for a confidence level of 95%. The expanded uncertainties along the line in Fig. 5(b) are shown in Fig. 5(c) as the red shadow area. The expanded uncertainty of the stray field, $U_{HZ}$, does not exceed 13.9 kA/m in the scanned area which is about 12.4 % of the maximal value of $H_z^S$.

It should be stressed that the reference sample can only represent stray field signals having a certain spatial frequency spectrum. As the TTF of the tip is determined based on the spectrum of the reference sample, it can only be meaningfully characterized within the spectrum as well. That is, to accurate measure a magnetic scale sample with microscale or submillimeter scale pole sizes, it requires that the spatial frequency spectrum of the test sample is covered by that of the reference sample.

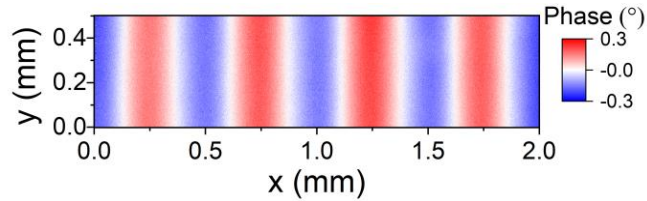

Fig. 6 MFM image of a magnetic scale sample with a lift height of $z = 150$ µm.

As an example, Fig. 6 shows a large range MFM image measured on a 2 mm-thick magnetic scale provided by SENSITEC GmbH, in which one surface was magnetized periodically in the out-of-plane direction with a pole size of about 250 µm. The scan size of 2 mm × 0.5 mm with a pixel size of 200 nm × 2.5 µm and the lift height of 150 µm are applied, accounting for the large domain size and the strong magnetic stray field. To determine the stray field of this sample based on the deconvolution process in the Fourier domain, the calibrated tip on S0 is not suitable anymore, due to the absence of the low spatial frequency contents of the TTF based on Fig.2(a). To solve this problem, an appropriate reference sample which exhibits a large domain size and with well-known magnetic properties must be applied to the tip calibration. After that, the presented quantitative MFM and its uncertainty evaluation process can be performed. This issue will be further studied and could be relevant for comparing quantitative MFM with other techniques for measuring stray field distributions based on, e.g. magneto optic imaging films, scanning hall sensors, and magnetic resistance sensors, and for bridging stray field measurements from a nanometer scale to a submillimeter scale.

## VI. Conclusion

In this work, we presented the magnetic tip calibration and stray field determination processes based on the deconvolution operation in Fourier domain in combination with the L-curve criterion for the regularization. Beginning from the functional relationships between the measurand and input quantities, we proposed an uncertainty evaluation process based on the GUM method. The quantitative MFM measurement and its uncertainty evaluation were exemplarily shown for the samples with nanometer scale domain sizes. For the chosen nanosensors PPP-MFMR tip, we estimated the expanded uncertainty $U_{TTF}$ at the tip apex to be about 17% of the maximal value of the *TTF*. In the stray field determination process, the expanded uncertainty $U_{HZ}$ is not more than 13.9 kA/m in the scanned area, about 12.4% of the maximal $H_z^S$ of the test sample on the nanometer scale (S1). The analysis of the uncertainty evaluation showed that the uncertainty component involving the deconvolution process for the tip calibration was one order smaller than other uncertainty components and can be ignored. In contrast, for the stray field determination it is comparable with the other uncertainty components due to the uncertainty of the calibrated tip. The presented uncertainty evaluation process could be principally applied to magnetic samples with domain sizes in a range from nanometer scale to submillimeter scale provided that an appropriate reference sample which covers the spatial frequency spectrum of the test sample is available. Quantitative stray field calibration process together with the uncertainty evaluation could be also applied to other techniques for determining the stray field, such as scanning hall magnetometry and magnetic optical indicator film technique, where a convolution between the sensor and the test sample has to be considered.

Acknowledgment

The work was funded within EMPIR JRP 15SIB06 NanoMag. The authors thank SENSITEC GmbH for providing the magnetic scale samples.